\documentclass[twocolumn,floatfix,footinbib,aps,prl,10pt,longbibliography]{revtex4-1}
\usepackage{amssymb}
\usepackage{graphicx}
\usepackage{amsmath}
\usepackage{color}
\usepackage{subfigure}
\usepackage{mathrsfs}
\usepackage{bm}
\usepackage{times}
\usepackage{amsfonts}
\usepackage{epstopdf}
\usepackage{enumerate}
\usepackage{soul}

\graphicspath{{img/}}

\begin{document}
\title{Finite Temperature Phase Behavior of Viral Capsids as Oriented Particle Shells}
\author{Amit R. Singh$^{1,2}$, Andrej Ko\v{s}mrlj$^{3,4}$, and Robijn Bruinsma$^{5}$}
\affiliation{$^{1}$Department of Physics and Astronomy, Johns Hopkins University, Baltimore, MD 21218, USA}
\affiliation{$^{2}$Currently at Department of Mechanical Engineering, Birla Institute of Technology and Science, Pilani, RJ 333031, India}
\affiliation{$^{3}$Department of Mechanical and Aerospace Engineering, Princeton University, Princeton, NJ 08540, USA}
\affiliation{$^{4}$Princeton Institute for the Science and Technology of Materials (PRISM), Princeton University, Princeton, NJ 08544, USA}
\affiliation{$^{5}$Departments of Physics and Astronomy, and Chemistry and Biochemistry, University of California, Los Angeles, CA 90095, USA}

\begin{abstract}
A general phase-plot is proposed for discrete particle shells that allows for
	thermal fluctuations of the shell geometry and of the inter-particle
	connectivities.  The phase plot contains a first-order melting
	transition, a buckling transition and a collapse transition and is used
	to interpret the thermodynamics of microbiological shells.
\end{abstract}
\maketitle

The development of shells that protect microbiological systems from a hostile
environment yet still allow for exchange of key nutrients was an essential step
in the evolution of life~\cite{alberts}. These shells are composed of molecules
decorated with both hydrophobic and hydrophilic groups (``amphiphiles'') in
such a way that in an aqueous environment they assemble into closed,
semi-permeable shells. An important example are the amphiphilic protein shells
that surround viruses~\cite{Baker1999} as well as many bacteria and most
archaea~\cite{sleytr}. Cryogenic-based microscopy studies~\cite{Baker1999} had
indicated that these ``capsids'' in general are strictly organized,
crystallographic structures (usually icosahedral or helical)~\cite{caspar} but
this view is being challenged. Thermodynamic studies of viral
self-assembly indicate, for the case of the assembly of viruses in solutions
with higher protein concentrations, that the interaction energies between
capsid proteins (``subunits'') should be no more than a few times the thermal
energy at room temperature in order to avoid the production of malformed
capsids~\cite{zlotnick1, zlotnick2, brooks}\footnote{This is due to kinetic
trapping. For lower protein concentrations, the probability of kinetic trapping
is reduced and interaction energies can be higher.~\cite{dykeman5361}}. Finite
temperature studies also showed that, due to thermal fluctuations, at least
some viral capsids are dynamical in nature and that the dynamics plays a role
in the life-cycle of the virus~\cite{bothner, speir2006}. Some capsids are even
in a molten or ``pleomorphic'' state~\cite{battisti},~\footnote{The Archaeal
and Bunyaviridae families of viruses (which includes the Hanta
viruses)~\cite{prang, overby, pietila} are important examples. The Gaussian
curvature of some of the Archaeal pleomorphs evolves over time~\cite{prang},
indicating that they are in a fluid-like state~\cite{perotti,vitelli}}.
Finally, all-atom Molecular Dynamics (MD) simulations of capsids revealed that
they can \textit{collapse} under the action of thermal
fluctuations~\cite{freddolino}.

The study of the melting and thermal collapse of a shell with a limited number
of constituent components ($10^2 - 10^3$  is an interesting statistical physics
problem in its own right. The geometry of the shell over which the components
are moving itself is defined by the position vectors of these same particles
and hence subject to thermal fluctuations~\cite{Peliti}. Here, we propose a
generic phase-diagram for the melting and collapse of discrete shells obtained
by comparing MD simulations of a coarse-grained model of capsids with the
continuum theory of thermally fluctuating surfaces. 

\begin{figure}
    \centering
    \includegraphics[width=3.4in]{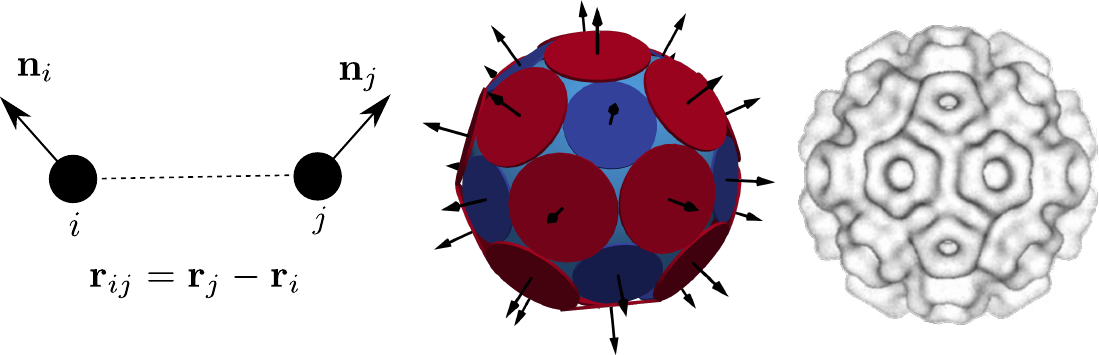}
    \caption{Left: Oriented particles $i$ and $j$ are separated by
    ${\bf{r}}_{ij}= {\bf{r}}_i - {\bf{r}}_j$ with orientations indicated by the
    unit vectors ${\bf{n}}_i$ and ${\bf{n}}_j$. Middle: The icosahedral
    groundstate configuration of an $N=32$ oriented particle system.
    The particle positions are displayed as the centers of close-packed disks. The
    orientations are displayed as the normals to the disks. Right:
    micrograph of the Cowpea Chlorotic Mottle Virus~\cite{liu} with 32
    capsomers.}
    \label{OPS}
\end{figure}

We first discuss the MD simulations. The coarse-grained model is based on the
so-called \textit{Oriented Particle System} (or ``OPS'')~\cite{OPS}. An OPS is
defined as a cluster of $N$ orientable and interacting point particles located
at ${\bf{r}}_i$. An orientation-dependent pair interaction $V({\bf{r}}_i,
{\bf{n}}_i ; {\bf{r}}_j, {\bf{n}}_j)$ acts between particle pairs \textit{i}
and \textit{j} with a separation vector ${\bf{r}}_{ij}= {\bf{r}}_i -
{\bf{r}}_j$ and unit vectors ${\bf{n}}_i$ and ${\bf{n}}_j$ describing their
orientations (see Fig.~\ref{OPS} (left)).
The oriented pair interaction used in the simulations was
\begin{equation}\label{ch3:H2}
\begin{split}
&V({\bf{r}}_i, {\bf{n}}_i ; {\bf{r}}_j, {\bf{n}}_j) = V_m
\left(1-e^{-\alpha\left(|{\bf{r}}_{ij}|-a\right)}\right)^2\\
&\quad \quad \quad \quad +K|{\bf{n}}_i
-{\bf{n}}_j|^2+
K\left(({\bf{n}}_i+{\bf{n}}_j)\cdot\widehat{{\bf{r}}}_{ij}\right)^2.
\end{split}
\end{equation}	
The first term is the Morse pair interaction~\cite{Morse}, with the binding
energy $V_m$, the equilibrium bond distance $a$, and the width of potential
well $1/\alpha$.  The second and third terms are known as the ``co-normality''
and ``co-circularity'' terms of an OPS system. Together, these two terms are
minimized if the two particles have the same orientation and if that shared
orientation is perpendicular to the unit vector $\widehat{{\bf{r}}}_{ij}$ that
is directed along the separation vector. Only interactions between particles
that are nearest neighbors are included, where the set of nearest neighbors can
change over time due to thermal fluctuations.

An OPS can be viewed as a coarse-grained representation of a viral capsid by
having the particle locations correspond to the centers of the ``capsomers'' of
viral capsids. The latter are disk-like groups of either six or five subunits
that frequently act as the basic building blocks of a
capsid~\cite{caspar,Speir}.  The orientational degrees of freedom correspond to
the normals to the capsomers, the depth $V_m$ of the Morse potential to the
capsomer binding energy (of the order of a few $k_BT$~\cite{Ceres, zlotnick}),
and the length scale $a$ to the diameter of a capsomer (of the order of a few
nanometer~\cite{Baker1999}).  Because the range of the hydrophobic attraction
between capsomers is short compared to their diameter, the dimensionless
parameter $\alpha a$ characterizing the width of the Morse potential needs to
be significantly larger than one. We used $\alpha a\approx4.621$. Next, $K$ is a
measure of the bending stiffness of the shell~(estimated to be in the range of
$10^2 k_BT$~\cite{Nguyen2005,Roos2010}). Finally, because a large energy
penalty is known to obstruct the removal of single capsomers from assembled
shells~\cite{Morozov2009}, evaporation of particles from the OPS shell is
suppressed by a soft fixed-area constraint imposed via the Augmented Lagrange
Multiplier method (see~\cite[Section I]{SupMat}). 

Figure~\ref{OPS} (middle) shows the minimum energy state of an $N=32$ OPS for
the case that $K/V_m$ is large compared to one. The shell has icosahedral
symmetry with the twelve blue disks indicating the five-fold symmetry sites of
the icosahedron (for actual capsids, these disks would correspond to pentameric
protein capsomers while the remaining twenty red disks would correspond to
hexameric capsomers). This structure should be compared to that of the ``T=3''
icosahedral pattern~\cite{caspar} of the $32$ capsomers of the capsid of the
Cowpea Chlorotic Mottle virus (CCMV)~\cite{liu} shown in the right of
Figure~\ref{OPS}.

\begin{figure}
    \centering
    \includegraphics[scale=0.5]{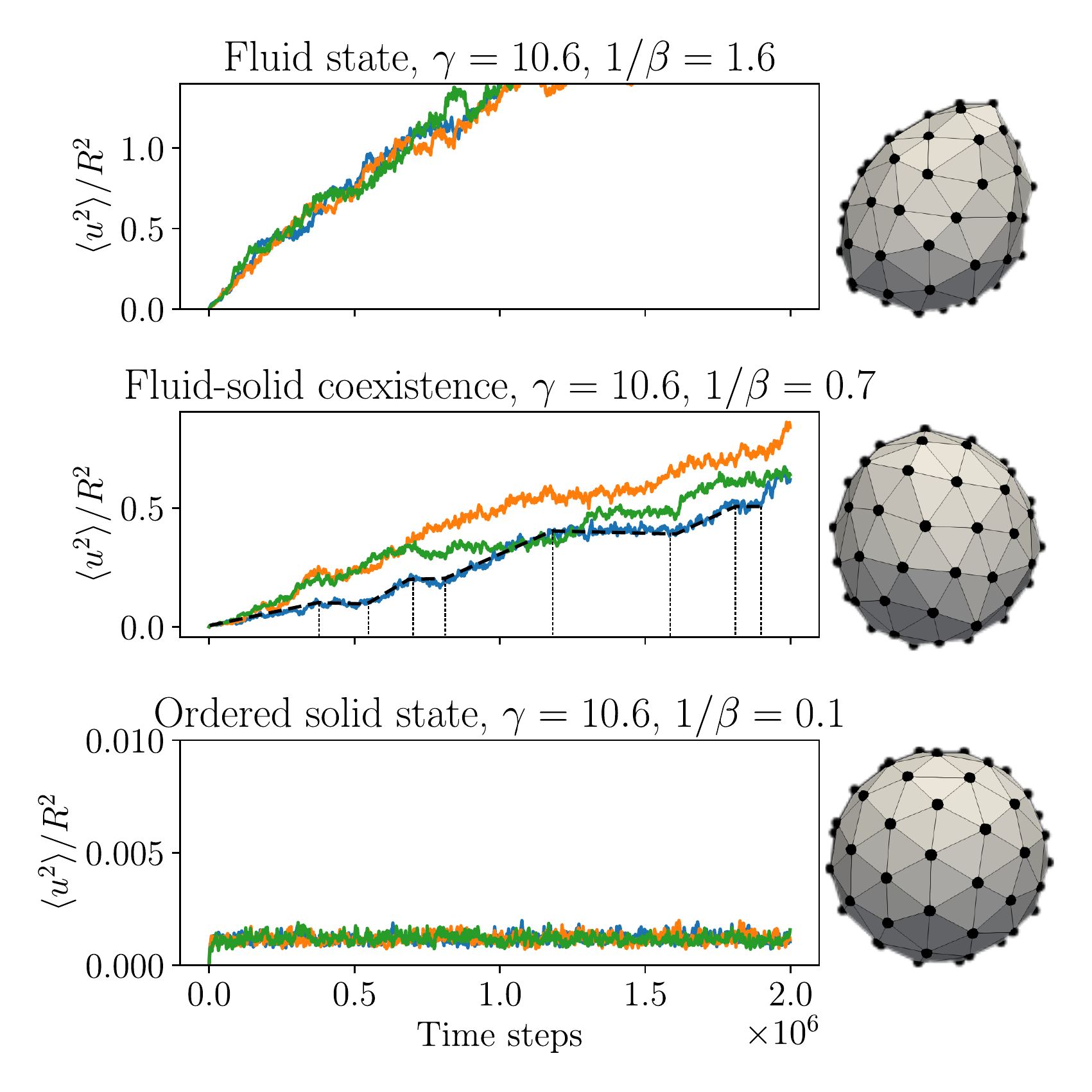}
    \caption{Plots of the mean square $\langle u^2(t) \rangle$ of the particle
    displacements (vertical axes) as a function of time (horizontal axis) for
    the FvK number $\gamma=10.6$ and three different temperatures $\beta^{-1}$.
    Three separate simulation runs are shown in each case (orange, blue, and green
    respectively). Bottom: for low temperature $\beta^{-1}=0.1$, $\langle u^2(t) \rangle$ reaches a
    constant value after a short transient (ordered solid state). The three runs are
    statistically similar. Top: for the higher temperature $\beta^{-1}=1.6$, $\langle u^2(t) \rangle$
    is proportional to time for the three runs, indicating diffusion (fluid state). The three
    runs are still similar. Middle: for intermediate temperature $\beta^{-1}=0.7$ the
    mean square displacements curves alternate between intervals where $\langle u^2(t) \rangle$
    steadily increases in time and intervals where it is roughly constant (fluid-solid coexistence). The
    blue curve is partitioned in this manner. There are large variations
    between different simulation runs. Snapshots on the
	right correspond to the final configurations for the simulation runs in
	blue color.}
    \label{DP}
\end{figure}

\begin{figure}
    \centering
    \includegraphics[scale=0.5]{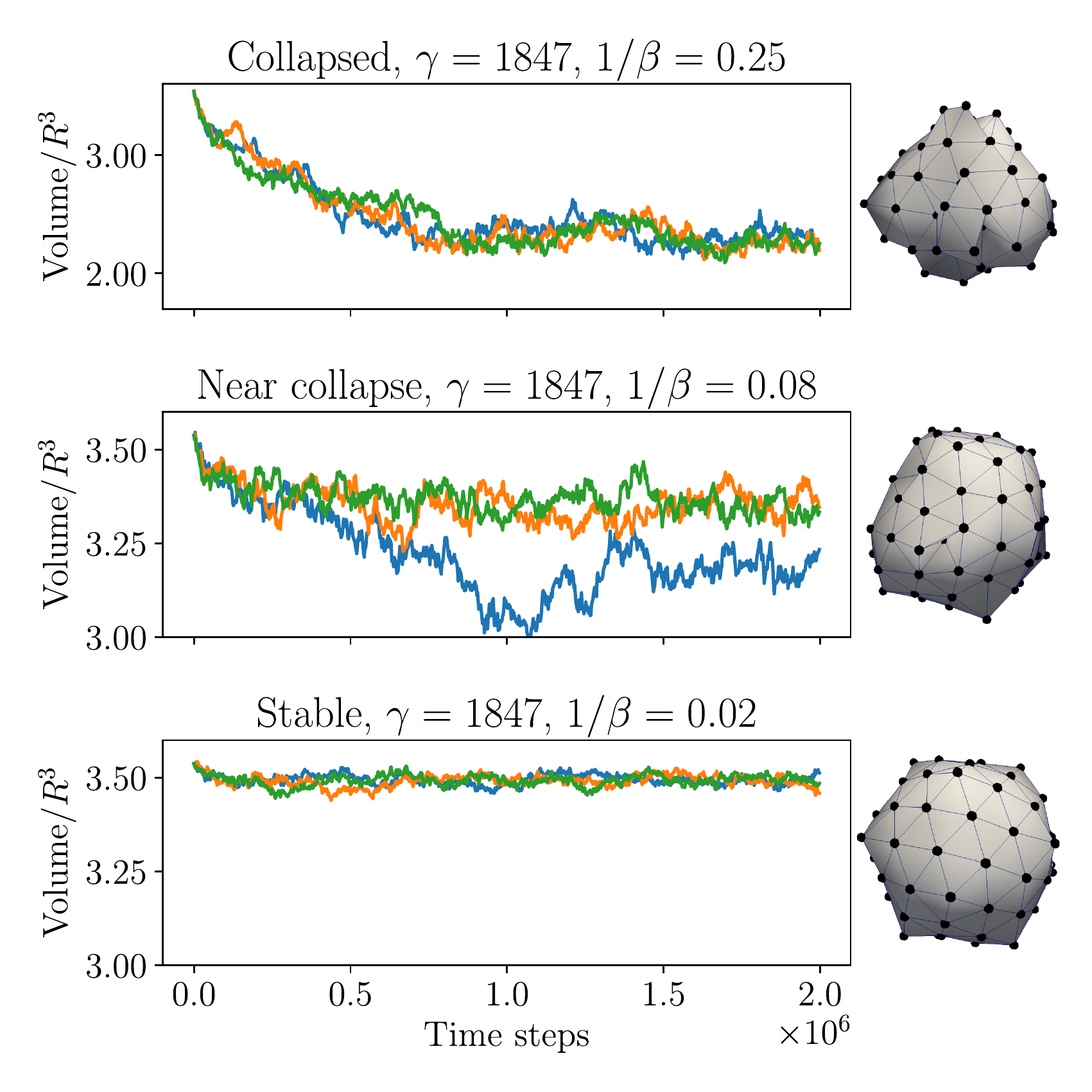}
    \caption{Time traces of the shell volume for the FvK number $\gamma=1847$
	and three different temperatures $\beta^{-1}$. Three separate
	simulation runs are shown in each case (orange, blue, and green
	respectively). Bottom: for low temperature $\beta^{-1}=0.02$ shell
	volumes reach a steady state after a small initial reduction. The three
	runs are statistically similar.  Top: for higher temperature
	$\beta^{-1}=0.25$ the shell volumes drastically decrease indicating the
	collapse of shells. The three runs are still similar. Middle: for
	$\beta^{-1}=0.08$ near the critical temperature for collapse, shell
	volumes exhibit large fluctuations and some shells undergo collapse
	(blue), while others remain stable (orange, green). Snapshots on the
	right correspond to the final configurations for the simulation runs in
	blue color.}
    \label{CT}
\end{figure}

Next, we carried out Brownian Dynamics simulations of $N=72$ OPS systems using
computational methods discussed further in the Supplemental
Materials~\cite[Section I]{SupMat}. The phase behavior was determined in terms
of the two thermodynamic parameters $\beta^{-1}=k_BT/V_m$, a dimensionless
measure of temperature in units of the depth $V_m$ of the pair interaction, and
$\gamma=2\alpha^2 V_mR^2/(3K)$ a dimensionless measure of the inverse of the
bending stiffness $K$ ($R\sim a N^{1/2}$ is the shell radius). In continuum
theory $\gamma$ is known as as the F\"oppl-von K\'arm\'an (FvK)
Number~\cite{Lidmar2003}. For different values of these two parameters we
encountered ordered, molten, buckled, and collapsing shells. Representative
realizations are shown in Figs.~\ref{DP} and~\ref{CT}.

The degree of fluidity of a shell was monitored using a dynamical method based
on plots of the mean square of the particle separations $\langle u^2(t)\rangle
\equiv\langle|{\bf{u}}_i(t)-{\bf{u}}_j(t)|^2\rangle$ as a function of time $t$,
averaged over all pairs $(i,j)$ of particles that were nearest neighbors in the
initial configuration~\cite{Maret}. For the present case, if in the long time
limit $\langle u^2(t)\rangle$ saturated (on average) to a constant value much
smaller than $R^2$ then the shell was assigned to be in a solid state. If, on
the other hand, $\langle u^2(t) \rangle$ increases linearly in time until it
reaches $ R^2$ -- which is consistent with particle diffusion -- then the shell
was assigned to be in a fluid state. Finally, when plots of $\langle u^2(t)
\rangle$ showed a random sequence of alternating time intervals of saturation
and linear growth for a given simulation run with drastic variations between
different runs then the shell was assigned to be in a fluid-solid coexistence
state. Examples of these three cases are shown in Fig.~\ref{DP} for
$\gamma=10.6$. In the low-temperature solid state, with $\beta^{-1}=0.1$, the
shell shape is spherical while in the coexistence regime, with
$\beta^{-1}=0.7$, significant shape-fluctuations are visible with
characteristic length-scales of the order of the shell radius. The particle
array still maintains local positional order but this has largely disappeared
for $\beta^{-1}=1.6$ (fluid state).

The collapse of shells is a pronounced feature of the phase behavior of shells
with larger values of the FvK Number $\gamma$. Collapse was monitored by
computing the volume of a continuous and differentiable surface that
interpolates between the particle locations, which was constructed using the
Loop shell subdivision method~\cite{Cirak,Feng}. Figure~\ref{CT} (top) shows an
example for $\gamma=1847$ of irreversible collapse induced by thermal
crumpling, as indicated by a drastic reduction in volume over time and the
production of very irregular shell shapes. In Fig.~\ref{CT} (bottom) this
simulation was repeated at a reduced temperature.  After a small initial
reduction, the shell volume reached a steady state~ \footnote{Note that shape
fluctuations always have to decrease the volume of shells, because the area of
shells was kept fixed}. The low-temperature shell shape is now
\textit{icosahedral}. In Fig.~\ref{CT} (middle) simulation were performed near
the critical temperature for collapse. In this case shell volumes exhibit large
fluctuations, with some shells undergoing a first order-like collapse
transition (blue time trace), while other shells remained stable over the
simulated time interval.

\begin{figure}
    \centering
    \includegraphics[scale=0.5]{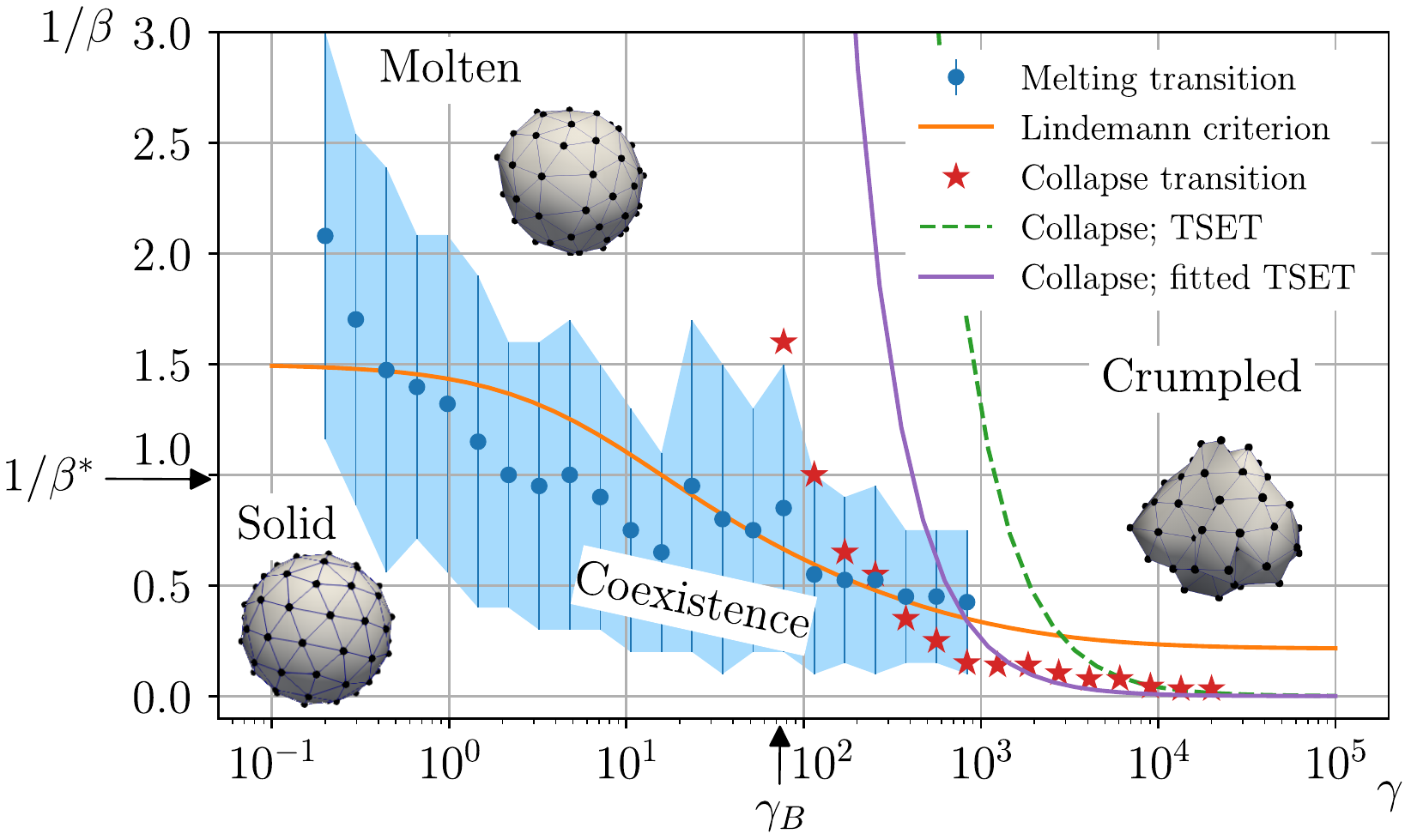}
    \caption{Phase plot of the $N=72$ oriented particle system. The vertical
	axis is the dimensionless temperature $\beta^{-1}$ and the horizontal
	axis the FvK number $\gamma$. The KTHNY transition temperature
	${\beta^*}^{-1}$ for melting of flat sheets and the buckling threshold
	$\gamma_B$, where the minimum energy state goes from spherical
	(Fig.~\ref{DP} (bottom)) to polyhedral (Fig.~\ref{CT} (bottom)), are
	marked by arrows. The vertical blue bars mark intervals of phase
	coexistence separating solid and molten states. The orange line marks
	the melting temperature obtained from a combination of the Lindemann
	criterion and thin-shell elasticity theory (TSET). The red stars mark
	the onset of irreversible collapse obtained from the simulations. The
	green dashed line marks the onset of irreversible crumpling/collapse
	predicted by TSET and the purple line is a fit of the TSET scaling
	relation with an overall numerical scale factor to account for
	discreteness effects.}
    \label{fig:PD}
\end{figure}

By collecting simulation runs for different values of the $\beta^{-1}$ and
$\gamma$ parameters, the phase plot of Fig.~\ref{fig:PD} was produced.
The vertical bars indicate temperature intervals over which fluid-solid
coexistence was observed following the criterion discussed above, with
the solid blue dots indicating midpoints. The large coexistence interval for
small $\gamma$ indicates that the melting transition should be first-order on a
rigid spherical surface. This is consistent with simulations of flat sheets of
particles interacting via Morse potential~\cite{Morse} for values of $\alpha a$
in the relevant range of $4.6$.  The onset of irreversible collapse of shells
for increasing temperature is indicated by orange triangles. Shells with
$\gamma \gtrsim 10^3$ collapsed before the particle array could melt. On the
other hand, shells with FvK numbers $\gamma$ in the range of $10^2-10^3$ would
melt with increasing temperature before they collapsed.

In order to interpret the phase plot, we compared it with the thin-shell
elasticity theory (TSET)~\cite{Seung} in which a curved and stretched layer is
assigned a bending and stretching energy given by
\begin{equation}
\begin{split}
\mathcal{H}=\int ds \, \frac{\kappa}{2}\,H^2 + \int ds\, \frac{1}{2}\,\left(\lambda {u_{ii}}^2+2\mu {u_{ij}}^2\right).
\end{split}
\end{equation}
Here, $\kappa$ is known as the Helfrich bending modulus of the layer, $H$ the
local mean curvature, $\mu$ and $\lambda$ are two-dimensional (2D) Lam\'{e}
coefficients, while $u_{ij}$ is the strain tensor. The FvK number equals
$\gamma=Y R^2/\kappa$ with $Y$ the corresponding 2D Young's modulus.  For a
flat sheet of point particles interacting via the Morse potential,
$\lambda=\mu$ while $Y=(8/3)\mu$. In terms of the parameters of the OPS
potential, $Y=4\alpha^2 V_m/\sqrt{3}$ and $\kappa=2\sqrt{3}K$. 

According to TSET, $\gamma$ determines the groundstate shape of a thin elastic
shell~\cite{Lidmar2003} such that for $\gamma$ less than a critical value
$\gamma_B$ (the ``buckling threshold'') that is in the range of $10^2-10^3$,
the shell has an approximately spherical shape (such as the spherical shape of
Fig.~\ref{DP} (bottom)) while for $\gamma$ above that threshold, the groundstate shape
is approximately polyhedral (such as the icosahedral shape of Fig.~\ref{CT} (bottom)).
As indicated in the phase plot, we find a buckling threshold around
$\gamma_B\simeq75$. The discrepancy between the computed and predicted values
of the buckling threshold, which has been noted before~\cite{singh}, is a first
measure of the importance of the discreteness effects. In order to apply TSET
to melting, one can combine it with the \textit{Lindemann Melting Criterion}
(LMC), which states that melting occurs when the RMS of the fluctuations of
particle displacements 
exceeds a certain fraction of the equilibrium interparticle spacing $a$. The LMC is known to work well for melting on flat two-dimensional surfaces~\cite{Maret}. In the regime
of harmonic fluctuations, the mean square $\langle \vec u\,^2\rangle$ of the
in-plane fluctuations and the mean square $\langle f^2\rangle$ of the
out-of-plane fluctuations can be shown to have the scaling form $\langle \vec
u\,^2\rangle = (k_BT/Y) G^u_{\infty}(\gamma)$ and $\langle f^2\rangle=(k_BT/Y)
G^f_{\infty}(\gamma)$, respectively. The scaling functions
$G^u_{\infty}(\gamma)$ and $G^f_{\infty}(\gamma)$ are discussed in the
Supplemental Materials~\cite[Section II]{SupMat}. Formally, TSET theory
corresponds to the limit of particle shells with $N$ large and $a$ small but
with fixed  $R\sim aN^{1/2}$. In order to include discreteness effects, we
expanded the three displacement fields in terms of a series of spherical
harmonics $\text{Y}_{\ell,m}$ and demanded that the total number of
out-of-plane modes $\mathcal{N}(\ell_\text{max})=(\ell_\text{max}+1)^2$ -- with
$\ell_\text{max}$ the maximum value of the quantum number $\ell$ -- equals the
number $N$ of particles minus two \footnote{The total number of degrees of
freedom is $3N-6$, because translation and rotation are prevented}.  For a
shell of $N=72$ particles $\ell_\text{max}=7$ and $\ell_\text{max}=8$ are
reasonable choices (since 72 is in the interval between $8^2$ and $9^2$). The
corrected scaling function has the same mathematical form as the TSET case but
it is larger by an overall constant scale factor in the range of 10-100,
depending on the value of $\ell_\text{max}$: discreteness thus strongly
amplifies the effects of thermal fluctuations. The resulting LMC melting
temperatures $T_m(\gamma) = T_m^0 G^u_N(0)/[G^u_N(\gamma)+G^f_N(\gamma)]$ are
plotted in Fig.~\ref{fig:PD} (orange line) with the melting temperature $T_m^0$
for $\gamma=0$ treated as a fitting parameter~\footnote{For small $\gamma$, the
scaling function $G^f_N(\gamma)$ is proportional to $\gamma$ while $G^u_N
(\gamma)$ is constant. The reduction of the melting temperature with increasing
$\gamma$ is thus dominated by the out-of-plane fluctuations.}. The resulting
fit is reasonable for the range of $\gamma$, where melting was observed.

According to TSET, elastic shells with large $\gamma$ should undergo a collapse
transition with increasing temperature ~\cite{paulose, kosmrlj,
baumgarten2018buckling}. Physically, this is due to the fact that crumpled
shells have a much larger configurational space for shape fluctuations than
(nearly) spherical shells. So their entropy is much larger as well, while the
volume of a crumpled shell with fixed area is correspondingly reduced. For
larger $\gamma$, the enthalpic cost of crumpling the surface is diminished so a
crumpling/collapse transition is to be expected.   The collapse of a shell
requires overcoming a free energy barrier that according to TSET vanishes when
$k_BT\gamma^{1/2}/\kappa$ is about $10^2$ (green dashed line in
Fig.~\ref{fig:PD} \cite{kosmrlj, baumgarten2018buckling}). For the
crumpling/collapse transition that we observed (red stars), solid shells (but
not liquid shells) roughly obeyed this scaling relation except that the value
of $k_BT\gamma^{1/2}/\kappa$ had to be decreased by a factor of about $10$
(purple line in Fig.~\ref{fig:PD}). We interpret this as a discreteness effect
similar to the one encountered for the melting transition. In the range of
$\gamma \sim 10^3 - 10^4$ the fitted thin-shell elasticity theory (purple line)
provides a reasonable estimate for the collapse transition. Small discrepancies
with the molecular dynamics simulation (red stars) are attributed to thermally
activated escape events over the energy barrier~\cite{baumgarten2018buckling}.
This is reflected in the middle panel of Fig.~\ref{CT}, where shell volumes
exhibit large fluctuations, with some shells undergoing a first order-like
collapse transition (blue time trace), while other shells remained stable over
the simulated time interval. In the range of $\gamma \sim 10^2-10^3$ there is a
much larger discrepancy between the thin-shell elasticity and molecular
dynamics simulations. For these values of $\gamma$ the collapse transition is
in the same range as the melting transition. The unbound dislocation pairs that
are forming during melting may significantly affect the collapse transition and
this effect is not captured in the TSET.

As an example how the phase plot Fig.~\ref{fig:PD} could be applied to viral
capsids, we compared the $N$=72 OPS with what is known about the phase
properties of viral capsids having 72 capsomers. It should be noted that our
phase plot corresponds to empty viral capsids, while the effect of osmotic
pressure due to the packaged DNA would increase both the melting and collapse
temperatures due to the suppression of radial fluctuations as discussed
in~\cite{kosmrlj}. In the Caspar-Klug classification of capsids, icosahedral
shells with 72 capsomers are known as ``T=7'' structures~\cite{caspar}.
Medically important T=7 viruses are the human polyoma and papilloma
un-envelopeded double-stranded DNA viruses, which both have a diameter of about
50~nm~\cite{belnap}. Because the polyoma and papilloma capsids are quite
spherical, the value of $\gamma$ should, for these two cases, lie below the
buckling threshold in Fig.~\ref{fig:PD}. In contrast, two forms of the capsid
of the T=7 thermostable DNA bacteriophage P23-45 with diameters of 66 and 82~nm
are, respectively, weakly and strongly polyhedral~\cite{bay}. This progression
of T=7 shapes straddling the buckling threshold could be understood within TSET
by noting that $\gamma$ scales as the square of the shell diameter. According
to Fig.~\ref{fig:PD}, in the relevant regime of $\beta^{-1}\simeq
1$~\cite{zlotnick1, zlotnick2}, capsids should be rather unstable in this part
of the phase plot: prone both to melting and collapse. In actuality these
viruses are known to be quite stable but this is because of a post-assembly
\textit{maturation process}, during which the capsid subunits are linked
together by covalent bonds. The initial procapsids of the T=7 viruses, which
are not yet bonded together, still could be thermodynamically unstable.
Interestingly, the assembly of the procapsids of phages typically takes place
on a preformed spherical \textit{scaffold} that is disassembled during
maturation~\cite{bay}. It is very suggestive that one purpose of the scaffold
is to prevent this instability.  This could be checked experimentally by a
study of the thermodynamic stability of self-assembling mutant T=7 empty shells
for which maturation and/or scaffold formation is blocked. Instability of the
procapsid of the related, but much larger, T=13 Herpes-Simplex virus, should be
even more pronounced.

\begin{acknowledgments}
Acknowledgements: We would like to thank Luigi Perotti, Jeff Eldredge and Bill
Gelbart for useful discussions and the UCLA Mechanical and Aerospace
Engineering Department for continued support.  This work was supported
by NSF through DMR Grants No.  1610384 and 1836404, and the CAREER
award DMR-1752100.
\end{acknowledgments}


%

\end{document}


\title{Finite Temperature Phase Behavior of Viral Capsids as Oriented Particle Shells: Supplementary Material}
\author{Amit Singh$^{1}$, Andrej Ko\v{s}mrlj$^{2,3}$, and Robijn Bruinsma$^{4}$}
\affiliation{$^{1}$Department of Physics and Astronomy, Johns Hopkins University, Baltimore, MD 21218, USA}
\affiliation{$^{2}$Department of Mechanical and Aerospace Engineering, Princeton University, Princeton, NJ 08540, USA}
\affiliation{$^{3}$Princeton Institute for the Science and Technology of Materials (PRISM), Princeton University, Princeton, NJ 08544, USA}
\affiliation{$^{4}$Departments of Physics and Astronomy, and Chemistry and Biochemistry, University of California, Los Angeles, CA 90095, USA}


\maketitle

\section{Numerical Methods.}

\subsection{Variational Method}
The finite temperature simulations of the OPS were based on an iterative solution of the coupled discretized Langevin equations
\begin{equation}
    \label{eq:rforce}
    \frac{\partial U^{n+1}}{\partial \mathbf{r}_i^{n+1}} + \frac{k_BT}{D}
    \frac{(\mathbf{r}_i^{n+1}-\mathbf{r}_i^{n})}{\Delta t} -
    k_BT\sqrt{\frac{2}{D\Delta t}}\boldsymbol{\xi}^{n+1}_{i}= 0,
\end{equation}
for the particle locations. At every time step, all particle locations and
orientations were updated in parallel. Here, $n$ is the time index with
discrete time step $\Delta t$, $U^{n+1}=\frac{1}{2}\sum_{i\neq
j}V({\bf{r}}_i^{n+1}, {\bf{n}}_i^{n+1} ; {\bf{r}}_j^{n+1}, {\bf{n}}_j^{n+1})$
the total potential energy associated with OPS pair interaction Eq.~(\ref{eq:rforce}) at time
step $n+1$. $D$ is the translational diffusion coefficient and the
$\boldsymbol{\xi}^{n}_{i}$ are a set of $3N$ Gaussian random variables with
variance one. The orientational degrees at time step $n+1$ were obtained by
demanding that the torques
\begin{equation}
\boldsymbol{\tau}_i^{n+1}\equiv\mathbf{n}_i^{n+1}\times\frac{\partial U^{n+1}}{\partial \mathbf{n}_i^{n+1}}=0
\label{torque}
\end{equation}
on the particle orientations vanished at every time step. Physically,
``integrating-out'' of the orientational degrees of freedom produces an
effective interaction between the remaining translational degrees of freedom
of the $N$ point particles. The effective interaction is no longer the sum of
radial pair interactions between neighboring point particles but now incudes
more complex 3-body and higher-order interactions and longer-range
interactions mediated by the orientational degrees of freedom.

At every time step, the set of $5N$ equations for the same number of unknowns
$\mathbf{r}_i^{n+1}$ and $\mathbf{n}_i^{n+1}$ was solved by numerical
minimization of the expression
\begin{equation}
    \label{eq:Functional1}
\begin{split}
    &I^{n+1}=U^{n+1}+ \left(\frac{k_BT}{2D\Delta t}\right)\sum_{i}
   {\left(\mathbf{r}_i^{n+1} -\mathbf{r}_i^{n}\right)^2}
    \\&- k_BT\sqrt{\frac{2}{D\Delta t}} \sum_{i}\boldsymbol{\xi}_i^{n+1}
    \cdot\left(\mathbf{r}_i^{n+1} - \mathbf{r}_i^{n}\right).
\end{split}
\end{equation}
where, as before, the superscripts $n+1$ and $n$ denote time-steps. The
minimization of $I^{n+1}$ with respect to $\mathbf{n}_i^{n+1}$ was of course
restricted to rotations, which leads to Eq.~(\ref{torque}). The successive
rotations were stored in the form of rotation vectors perpendicular to the
orientation vectors. The orientations were reconstructed from the rotation
vectors using the method of quaternions~\cite{shoemake1985animating}.

\subsection{Area Constraint}

In order to suppress evaporation of single particles from the cell we imposed
a soft fixed area constraint using the Augmented-Lagrangian (AL) technique as
follows. The constrained minimization problem is finding the minimum of
$I^{n+1}$ subject to the constraint $A\left(\mathbf{r}^{n+1}\right) - A_0 =
0$ where $A_0$ is the zero-temperature area and
$A\left(\mathbf{r}^{(n+1)}\right)\equiv A^{n+1}$ is the area after $n + 1$ time steps.
Introduce a Lagrange multiplier term and an augmenting penalty term as
follows
\begin{equation}
\begin{split}
&F[\mathbf{r}^{(n+1)},\mathbf{n}^{(n+1)}] = I^{n+1} +
    \frac{k^{n+1}}{2}\left(A^{n+1} - A_0\right)^2 -
    \lambda^{n+1}\left(A^{n+1} - A_0\right)
\end{split}
\end{equation}%
where $k^{n+1}$ is an estimate of the spring constant of the penalty term at time step $n+1$
and $\lambda^{n+1}$ is an estimate of the Lagrange multiplier at time step $n+1$. For a given time step,
successive estimates are updated according to
\begin{enumerate}
    \item Set $k^{(n+1)} = 1000.0$ and $\lambda^{(n+1)} = 10.0$.
    \item Find $\mathbf{r}^{(n+1)}, \mathbf{n}^{(n+1)} = \mathrm{argmin}\
    F[\mathbf{r}^{(n+1)},\mathbf{n}^{(n+1)}]$
    \item While $\left(A^{(n+1)} - A_0\right) > 10^{-8}$, repeat
    \begin{enumerate}
        \item $\lambda^{(n+1)} \gets \lambda^{(n+1)} -
        k^{(n+1)}\left(A^{(n+1)} - A_0\right)$
        \item $k^{(n+1)} \gets 10\times k^{(n+1)}$
        \item Find $\mathbf{r}^{(n+1)}, \mathbf{n}^{(n+1)} = \mathrm{argmin}\
        F[\mathbf{r}^{(n+1)},\mathbf{n}^{(n+1)}]$
    \end{enumerate}
\end{enumerate}
At the end of this procedure the area constraint is satisfied to a desired
tolerance. The advantage of Augmented Lagrangian method over the standard
method of Lagrange multipliers is that one does not need to introduce an
extra degree of freedom. The Augmented Lagrangian parameters $k$ and
$\lambda$ are solved iteratively in a loop external to the regular time step
updating. The parameter $k^{(n+1)}$ does not need to go to infinity and thus
numerical ill-conditioning is avoided.

\subsection{Simulation Details}

Equation~(\ref{eq:Functional1}) can be written in non-dimensionalized form as
\begin{equation}
    \label{eq:Functional2}
    \begin{split}
    I[\mathbf{r}^{(n+1)},\mathbf{n}^{(n+1)}] =
    \sum_{i\neq j} \left( e^{-2\alpha\left(|\mathbf{r}_{ij}|-a\right)} -
    2e^{-\alpha\left(|\mathbf{r}_{ij}|-a\right)} \right)\\[0.25cm]
    + \frac{2\alpha^2R^2}{3\gamma}\left(\sum_{i\neq j}	|\mathbf{n}_i
    - \mathbf{n}_j|^2 + \sum_{i\neq j} \left( \frac{
        \left( \mathbf{n}_i + \mathbf{n}_j \right)
    \cdot\mathbf{r}_{ij}} {|\mathbf{r}_{ij}|}\right)^2\right)\\[0.25cm]
    + \frac{\zeta}{a^2}\frac{\left(\mathbf{r}^{(n+1)}-
\mathbf{r}^{(n)}\right)^2}{2} - \frac{1}{a}\sqrt{\frac{2\zeta}{\beta}}\,
\hat{\boldsymbol{\xi}_r}\cdot \left(\mathbf{r}^{(n+1)} - \mathbf{r}^{(n)}
\right)
\end{split}
\end{equation}
$V_M$, $\alpha$ and $a$ are parameters of the Morse potential that control the
depth, the width and the equilibrium separation respectively. $R$ is the
radius of the shell. $K$ controls the strength of the orientational
potentials. $\mu$ is the mobility. The three non-dimensional parameters are given as
\begin{equation}
    \begin{split}
    \label{eq:dimConst}
    \beta &= \frac{V_M}{k_BT},\\[0.25cm]
    \gamma &= \frac{2\alpha^2R^2V_M}{3K},\\[0.25cm]
    \zeta &= \frac{a^2}{V_M}\frac{1}{\mu\Delta t}.
    \end{split}
\end{equation}
The interactions between the particles are restricted to the nearest neighbors
that were obtained via a triangulation of the shell surface by first
projecting all particles to a sphere and then constructing the convex hull of
the particle array. The set of nearest neighbors consists of particles that
share an edge in the triangulation.

The width of the Morse potential can be expressed as $\delta =
\ln{2}/(\alpha a)$ and  we previously demonstrated that it must be larger than
a critical value for the formation of stable shells at low FvK
numbers.~\cite{singh2018study} In practice, we found that $\delta = 0.15$ was
sufficiently large and thus we chose $\alpha a = \ln{2}/0.15 = 4.621$.

We chose 30 log-spaced values of FvK number ($\gamma$) between 0.2 and 20000.
For each FvK number, we used 20 temperature ($1/\beta$) values. $\zeta$ is
kept fixed at $2.5\times10^5$. For each combination of FvK number and temperature
we did 3 runs where each run comprised of evolving the system for $2 \times
10^6$ time steps starting from the zero-temperature minimum energy structure
for the specific FvK number. After every time-step we ``subtract'' off the
rigid body translation and rotation with respect to the initial structure
using the Kabsch algorithm~\cite{Kabsch12999}.

\subsection{Simulation Output}

At every time step, we stored the means of the squared relative
neighbor-neighbor displacement for all particles, which is calculated as follows.
Suppose that particles $i$ and $j$ were the nearest neighbors at time $t=0$, then
the squared relative displacement is given by $\lVert\left(\mathbf{r}_i(t) - \mathbf{r}_i(0)\right) - \left(\mathbf{r}_j(t) - \mathbf{r}_j(0)\right)\rVert^2$. We cannot use displacement of individual particles $\lVert \mathbf{r}_i(t) -
\mathbf{r}_i(0)\rVert^2$ because for 2D melting it
diverges and relative neighbor-neighbor displacement provides the appropriate
modification to the Lindemann criterion~\cite{bedanov1985modified}.

We also stored, the asphericity, the volume and the root mean-squared angle
deficit of the shell. The asphericity is defined as $\langle (R_i -
\langle R_i \rangle)^2 \rangle/\langle R_i \rangle^2 $ where $R_i$ is the
radial distance of particle $i$ from center of the shell. For calculating
volume and root-mean-squared angle deficit, we need a triangulation of the
surface of the shell defined by the particles. We calculate the triangulation
by projecting each particle to a unit sphere and calculating the convex hull
of the spherical point cloud using CGAL~\cite{cgal:eb-19b} software package.
The angle deficit is a measure of Gaussian curvature of the surface and it is
calculated as discussed in~\cite{borrelli2003angular}.

We also store the particle positions and orientations after every 2000 time
steps. We use this to reconstruct the shell shape during post-processing to
detect crumpling of the shells.

\subsection{Software}

The minimizations were carried out using the Limited Memory
BFGS~\cite{zhu1997algorithm} algorithm and the code used is available
publicly on \url{https://github.com/amit112amit/ops-python}, in the form of a
Python wrapped C++ code. The main driver used for the simulations is
\texttt{phasediagramsimulation.py} located at
\url{https://github.com/amit112amit/ops-python/blob/master/phasediagramsimulation.py}.
The results of these simulations are publicly available as an interactive
Jupyter notebook at
\url{https://mybinder.org/v2/gh/amit112amit/opsresults/master?filepath=ShowPlots.ipynb}.

\section{Thin-Shell Elasticity Theory}
In this section we discuss how to calculate the spectrum of in-plane and out-of-plane fluctuations within the harmonic regime of the thin-shell elasticity theory. First, we discus the continuum limit, where the radius $R$ of the spherical shell is assumed to be much larger than its thickness $t$. Second, we discuss how to take into account finite size effects.

\subsection{Large shell limit}

As shown previously in refs.~\cite{paulose,kosmrlj}, the relevant length scale for the statistical mechanics of thin shells is the elastic length scale $\ell_\text{el}=R \gamma^{-1/4}\sim \sqrt{R t}$, where $\gamma = YR^2/\kappa$ is the FvK number. ($Y$ is the Young's modulus, $\kappa$ is the bending rigidity). For thin shells $\gamma \gg 1$ and thus $\ell_\text{el} \ll R$. In this limit it is sufficient to consider a small square patch of spherical shell, which is much larger than $\ell_\text{el}$ and much smaller than $R$~\cite{paulose,kosmrlj}.

Deformation of a small spherical patch is described with displacement vector fields, which are decomposed into the outward radial displacement field $f({\bf x})$ and the tangential displacements $u_i({\bf x})$, where ${\bf x}=(x_1,x_2)$ and $i\in\{1,2\}$.  The total deformation energy of a small patch consists of the bending energy cost
\begin{eqnarray}
U_\text{b} = \int \! dA \ \frac{1}{2} \kappa \left(\Delta  f\right)^2,
\end{eqnarray}
where $\kappa$ is the bending rigidity, and the stretching energy cost
\begin{eqnarray}
U_\text{s} = \int \! dA \left[\frac{1}{2} \lambda u_{ii}^2 + \mu u_{ij}^2 \right],
\end{eqnarray}
where $\lambda$ and $\mu$ are Lame elastic constants with the Young's modulus $Y=4 \mu (\mu+\lambda)/(2 \mu + \lambda)$ and the summation over repeated indices is implied. Here, we introduced the strain tensor
\begin{equation}
u_{ij}=\frac{1}{2} \left(\partial_i u_j + \partial_j u_i \right) + \delta_{ij}\ \frac{f}{R},
\end{equation}
where $\delta_{ij}$ is the Kronecker delta. Since we are only focusing on the harmonic spectrum of fluctuations, we neglected the nonlinear term $(\partial_i f)(\partial_j f)/2$ in the strain tensor $u_{ij}$, which becomes relevant, when the amplitude of fluctuations becomes larger than the shell thickness~\cite{kosmrlj}.

The spectrum of fluctuations can be analyzed with the help of Fourier transforms $f({\bf x}) = \sum_{\bf q} f({\bf q}) e^{i {\bf q} \cdot {\bf x}}$ and $u_i({\bf x}) = \sum_{\bf q} u_i({\bf q}) e^{i {\bf q} \cdot {\bf x}}$. Furthermore, we use the Helmholtz decomposition for the in-plane displacements ${\bf u}({\bf q})={\bf u}_\parallel({\bf q})+{\bf u}_\perp({\bf q})$, where ${\bf u}_\parallel \parallel {\bf q}$ and ${\bf u}_\perp \perp {\bf q}$. Using this decomposition we rewrite the total deformation energy as
\begin{eqnarray}
U_\text{b} + U_\text{s} &=& A \sum_{\bf q} \left(\frac{1}{2} \kappa q^4 |f({\bf q})|^2 + \frac{1}{2}(2 \mu+\lambda) q^2 |{\bf u}_\parallel({\bf q})|^2 + \frac{1}{2} \mu q^2 |{\bf u}_\perp({\bf q})|^2 + 2 \frac{(\mu+\lambda)}{R^2} |f({\bf q})|^2 \right.\nonumber \\
&& \left. \quad \quad \quad \ 
+i \frac{(\mu+\lambda)}{R} \big[{\bf q} \cdot {\bf u}_\parallel({\bf q})\big]f(-{\bf q}) - i \frac{(\mu+\lambda)}{R} \big[{\bf q} \cdot {\bf u}_\parallel(-{\bf q})\big]f({\bf q}) )
\right).
\end{eqnarray}
The spectrum of thermal fluctuations is thus 
\begin{subequations}
\begin{align}
\left<|f({\bf q})|^2 \right> & = \frac{k_B T}{A (\kappa q^4 + Y/R^2)}, \\
\left<|{\bf u}_\perp({\bf q})|^2 \right> & = \frac{k_B T}{A \mu q^2},\\
\left<|{\bf u}_\parallel({\bf q})|^2 \right> & = \frac{k_B T (4 (\mu+\lambda) + \kappa q^4 R^2)}{A (Y q^2 + \kappa q^6 R^2)(2 \mu+\lambda)},
\end{align}
\end{subequations}
where $T$ is temperature, $k_B$ the Boltzmann constant, and $A$ the area of the small spherical patch. The total amplitude of out-of-plane fluctuations can then be obtained as
\begin{subequations}
\begin{align}
\left< f({\bf x})^2 \right> &= \sum_{\bf q} \left<|f({\bf q})|^2 \right> \approx A \int \! \frac{d^2 {\bf q}}{(2 \pi)^2} \left<|f({\bf q})|^2 \right> \approx A \int_{\pi/R}^{\pi/a} \frac{q dq}{(2 \pi)} \left<|f(q)|^2 \right> \equiv \frac{k_B T}{Y} G^f_\infty(\gamma), \\
G^f_\infty(\gamma) & \approx  \left\{
\begin{array}{ll}
\sqrt{\gamma}/(2 \pi), & a \ll \ell_{el} \ll R\\
(\pi/4) (R/a)^2, & \ell_{el} \ll a \ll R\\
\end{array}
\right.,
\end{align}
 \label{eq:GF}%
\end{subequations}
where $a$ is the microscopic cutoff related to the interparticle spacing. Similarly, we calculate the total amplitude of in-plane fluctuations as
\begin{subequations}
\begin{align}
\left< {\bf u}({\bf x})^2 \right> &= \sum_{\bf q} \left(\left<|{\bf u}_\parallel({\bf q})|^2 \right> + \left<|{\bf u}_\perp({\bf q})|^2 \right> \right) \approx A \int_{\pi/R}^{\pi/a} \frac{q dq}{(2 \pi)} \left(\left<|{\bf u}_\parallel({q})|^2 \right> + \left<|{\bf u}_\perp({q})|^2 \right> \right) \equiv \frac{k_B T}{Y} G^u_\infty(\gamma), \\
G^u_\infty(\gamma) & \approx  \left\{
\begin{array}{ll}
(2/9\pi ) \left[\ln (\gamma) + 8 \ln(R/a)\right], & a \ll \ell_{el} \ll R\\
(8/3\pi) \ln(R/a) & \ell_{el} \ll a \ll R\\
\end{array}
\right.,
\end{align}
 \label{eq:GU}
\end{subequations}
where we used $\mu=\lambda=3Y/8$ that corresponds to the continuum limit of the OPS.

\subsection{Finite size effects}

In order to capture the finite size effects for small shells composed of $N$ particles, we have to consider deformations of the whole spherical shell, which are decomposed into the outward radial displacement field $f(\theta,\phi)$ and the tangential displacements $u_\alpha(\theta,\phi)$, where $\alpha\in\{\theta,\phi\}$. The Helmholtz decomposition is used for tangential displacements to separate the irrotational and the solenoidal part as~\cite{Zhang}
\begin{equation}
u_{\alpha} = D_\alpha \psi + \gamma_{\alpha \beta} D^\beta \chi.
\end{equation}
Here $D_\alpha$ are covariant derivatives and $\gamma_{\alpha \beta}$ is the alternating tensor, which depends on the metric and can be expressed as $\gamma_{\alpha \beta}=\sqrt{g} \epsilon_{\alpha \beta}$, where $g$ is the determinant of the metric tensor $g_{\alpha \beta}$ and $\epsilon_{\alpha \beta}$ is the antisymmetric Levi-Civita symbol. Indices are raised and lowered with the metric tensor $g_{\alpha \beta}$. The tangential displacements can thus be described with two fields $\psi(\theta,\phi)$ and $\chi(\theta,\phi)$. The 3 scalar fields describing displacements can be expanded in spherical harmonics as
\begin{eqnarray}
f(\theta, \phi) &=& r_0+\sum_{\ell=2}^{\ell_\text{max}}\sum_{m=-\ell}^\ell a_{\ell,m} \ R\  \text{Y}_{\ell,m} (\theta, \phi), \nonumber \\
\psi(\theta, \phi) &=& \sum_{\ell=2}^{\ell_\text{max}}\sum_{m=-\ell}^\ell b_{\ell,m} \ R^2\  \text{Y}_{\ell,m} (\theta, \phi), \nonumber \\
\chi(\theta, \phi) &=& \sum_{\ell=2}^{\ell_\text{max}}\sum_{m=-\ell}^\ell c_{\ell,m}\ R^2\  \text{Y}_{\ell,m} (\theta, \phi).
\end{eqnarray}
Note that we excluded spherical harmonics with $\ell=1$ that generate translations. The radial shrinking of shell $r_0$ is obtained from the fixed area constraint as
\begin{eqnarray}
r_0  &=& - \frac{R}{16 \pi} \sum_{\ell=1}^{\ell_\text{max}}\sum_{m=-\ell}^\ell (\ell^2 + \ell+ 2) |a_{\ell,m}|^2.
\end{eqnarray}
The cutoff $\ell_\text{max}$ is determined by requiring that the total number of degree of freedoms $3 (\ell_\text{max}+1)^2$ is equal to $3N-6$, where the 6 degrees of freedom are subtracted to prevent translations and translations. For a shell with $N=72$ particles we consider $\ell_\text{max}=7$ and $\ell_\text{max}=8$.

The total deformation energy can be rewritten as~\cite{Zhang}
\begin{eqnarray}
U_\textrm{b} + U_\textrm{s} &=& \int \! dA \left[\frac{\kappa}{2}  \left(\Delta  f + \frac{2 f}{R^2} \right)^2 + 2 (\mu+\lambda) \frac{f^2}{R^2} + 2 (\mu+\lambda)  \left(\Delta \psi \right)\, \frac{f}{R} \right. \nonumber\\
&& \quad  \quad \quad  \left. + \frac{(2 \mu + \lambda)}{2}   \left(\Delta \psi \right)^2  + \mu \frac{ \psi \left(\Delta \psi \right)}{R^2} + \frac{\mu}{2} \left(\Delta \chi \right)^2 + \mu \frac{ \chi \left(\Delta \chi \right)}{R^2} \right]. \nonumber \\
U_\textrm{b} + U_\textrm{s} &=&\sum_{\ell=1}^{\ell_\text{max}}\sum_{m=-\ell}^\ell \bigg[ \left(\frac{\kappa}{2} (\ell+2)^2 (\ell-1)^2  + 2 (\mu+\lambda) R^2 \right) |a_{\ell,m}|^2  - (\mu+\lambda) R^2 \, \ell (\ell+1) \left(a_{\ell,m} b^*_{\ell,m} + a^*_{\ell,m} b_{\ell,m} \right) \nonumber \\
&& \quad \quad \quad \quad \quad   + \frac{R^2\ell (\ell+1)}{2} \left[(2 \mu+\lambda) \ell (\ell+1) - 2 \mu \right]  |b_{\ell,m}|^2 + \frac{\mu R^2}{2} (\ell-1) \ell (\ell+1) (\ell+2)  |c_{\ell,m}|^2  \bigg].
\end{eqnarray}
The spectrum of fluctuations is thus
\begin{eqnarray}
\left<  |a_{\ell,m}|^2 \right> & = & \frac{k_B T}{\kappa (\ell+2)^2 (\ell-1)^2 + Y R^2  \frac{(\ell+2) (\ell-1)}{\ell (\ell+1) - 2 \mu / (2 \mu + \lambda)}}, \nonumber \\
\left<  |b_{\ell,m}|^2 \right> & = & \frac{k_B T}{(2 \mu +\lambda) R^2 \ell^2 (\ell+1)^2 - 2 \mu R^2 \ell (\ell+1) - \frac{4 (\mu+\lambda)^2 R^4  \ell^2 (\ell+1)^2}{[\kappa (\ell+2)^2 (\ell-1)^2 + 4 (\mu+\lambda) R^2]} } , \nonumber\\
\left<  |c_{\ell,m}|^2 \right> & = & \frac{k_B T}{\mu R^2 (\ell-1) \ell (\ell+1) (\ell+2)}.
\end{eqnarray}
The variance of radial fluctuations is 
\begin{eqnarray}
\left< \delta f^2 \right> & = & \sum_{\ell=2}^{\ell_\text{max}}\sum_{m=-\ell}^\ell R^2 \left< |a_{\ell,m}|^2 \right> \equiv  \frac{k_B T}{Y} G^f_N(\gamma), \nonumber\\
G^f_N(\gamma) & = & \sum_{\ell=2}^{\ell_\text{max}}\sum_{m=-\ell}^\ell \frac{\gamma}{(\ell+2)^2 (\ell-1)^2 + \gamma  \frac{(\ell+2) (\ell-1)}{\ell (\ell+1) - 2 /3}},
\label{eq:GFN}
\end{eqnarray}
where we used $\mu=\lambda=3Y/8$ that corresponds to the continuum limit of the OPS. Similarly, we calculate the variance of tangential fluctuations as
\begin{eqnarray}
\left< {\bf u}^2  \right> & = &  \sum_{\ell=1}^{\ell_\text{max}}\sum_{m=-\ell}^\ell R^2 \, \ell (\ell+1) \left[ \left< |b_{\ell,m}|^2  \right> + \left< |c_{\ell,m}|^2 \right> \right] \equiv  \frac{k_B T}{Y} G^u_N(\gamma), \nonumber \\
 G^u_N(\gamma) & = &   \sum_{\ell=2}^{\ell_\text{max}}\sum_{m=-\ell}^\ell  \left( 
\frac{8}{3 \left[3 \ell^2 (\ell+1)^2 - 2 \ell (\ell+1) - \frac{6 \gamma \ell^2 (\ell+1)^2}{(3 \gamma + (\ell+2)^2 (\ell-1)^2)}\right]}
 + \frac{8}{3 (\ell-1) \ell (\ell+1) (\ell+2)}
 \right)
 \label{eq:GUN}
\end{eqnarray}

\begin{figure}[h]
	\centering
	\includegraphics[scale=.33]{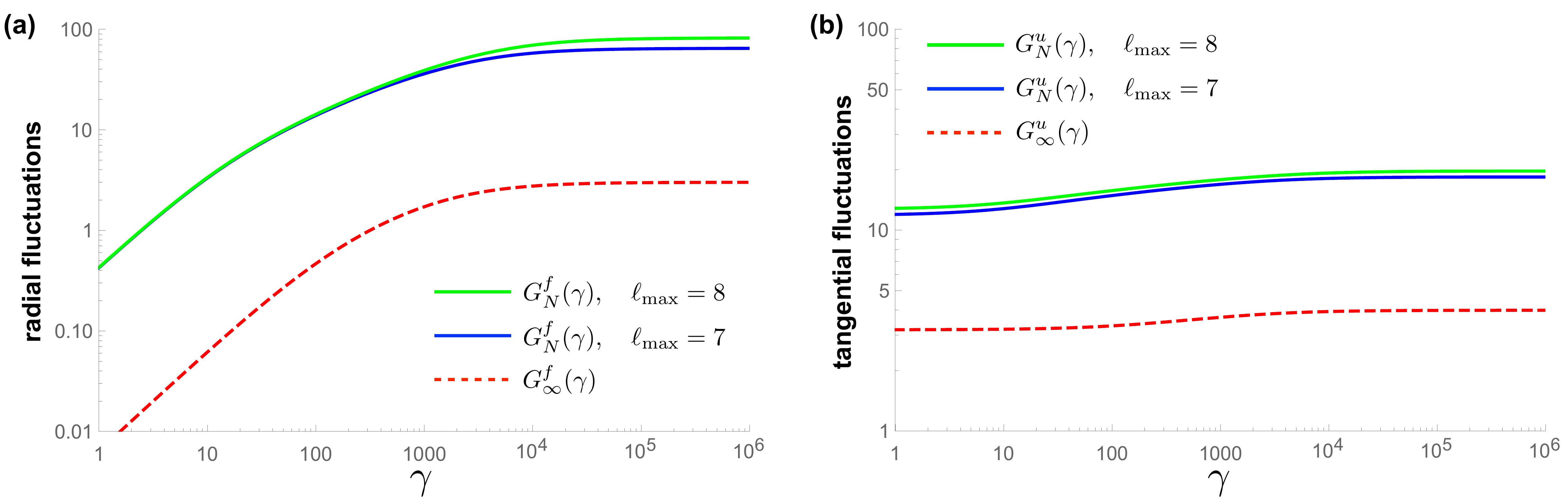}
	\caption{Comparison of (a) radial fluctuations and (b) tangential fluctuations between the continuum theory for large thin shells (red dashed lines, $R/a=2.2$) and finite size shells (blue, $\ell_\text{max}=7$, and green, $\ell_\text{max}=8$).}
	\label{FigS1}
\end{figure}

In Fig.~\ref{FigS1} we compare the scaling functions for radial displacements $G^f_N(\gamma)$ in Eq.~(\ref{eq:GFN}) and tangential displacements $G^u_N(\gamma)$ in Eq.~(\ref{eq:GUN}) for a shell with $N=72$ particles ($\ell_\text{max}=7$-$8$) to the ones obtained in the large shell limit ($G^f_\infty(\gamma)$ and $G^f_\infty(\gamma)$ in Eqs.~(\ref{eq:GF}) and (\ref{eq:GU}) with $R/a=2.2$). 
The radius $R=2.2 a$ was chosen, such that the area of the sphere is equal to the area of 140 equilateral triangles with side length $a$ that are covering the surface of the shell with $N=72$ particles, i.e. $4 \pi R^2 = 140 a^2 \sqrt{3}/4$. Because the shell radius is quite small, we didn't use asymptotic expressions in Eqs.~(\ref{eq:GF}b) and (\ref{eq:GU}b), but we numerically integrated expressions in Eqs.~(\ref{eq:GF}a) and (\ref{eq:GU}a).

%